\documentclass[prl,twocolumn,showpacs,superscriptaddress]{revtex4}

\usepackage{graphicx}
\usepackage{amsmath}
\usepackage{amssymb}

\begin{document}
\title{Entanglement quantification with Fisher information}
\author{Akira Kitagawa}
\email{kitagawa@nict.go.jp}
\affiliation{National Institute of Information and Communications
Technology (NICT) \\4-2-1 Nukui-Kita, Koganei, Tokyo 184-8795 Japan}
\affiliation{CREST, Japan Science and Technology Agency, 
1-9-9 Yaesu, Chuoh, Tokyo 103-0028 Japan}
\author{Masahiro Takeoka}
\affiliation{National Institute of Information and Communications
Technology (NICT) \\4-2-1 Nukui-Kita, Koganei, Tokyo 184-8795 Japan}
\affiliation{CREST, Japan Science and Technology Agency, 
1-9-9 Yaesu, Chuoh, Tokyo 103-0028 Japan}
\author{Masahide Sasaki}
\affiliation{National Institute of Information and Communications
Technology (NICT) \\4-2-1 Nukui-Kita, Koganei, Tokyo 184-8795 Japan}
\affiliation{CREST, Japan Science and Technology Agency, 
1-9-9 Yaesu, Chuoh, Tokyo 103-0028 Japan}
\author{Anthony Chefles}
\affiliation{Hewlett Packard Labs, 
Filton Road Stoke Gifford Bristol BS34 8QZ, UK}
\date{\today}
\begin{abstract}
\vspace{0.5cm}

We show that the Fisher information associated with entanglement-assisted 
coding has a monotonic relationship with the logarithmic negativity, 
an important entanglement measure, 
for certain classes of continuous variable (CV) quantum states 
of practical significance. These are the two-mode squeezed states 
and 
the non-Gaussian states obtained from them by photon subtraction. 
This monotonic correspondence can be expressed analytically 
in the case of pure states. 
Numerical analysis shows that this relationship holds to a very good approximation 
even in the mixed state case of  
the photon-subtracted squeezed states. 
The Fisher information is evaluated 
by the CV Bell measurement 
in the limit of weak signal modulation. 
Our results suggest that 
the logarithmic negativity of certain sets of 
non-Gaussian mixed states 
can be experimentally accessed 
without homodyne tomography, 
leading to significant simplification of 
the experimental procedure.

\end{abstract}


\maketitle

Some important quantum information processing protocols 
with continuous variable (CV) systems 
essentially require highly nonlinear processes 
beyond Gaussian operations. 
An important example is entanglement distillation 
for a two-mode squeezed state, 
which is impossible to perform with Gaussian local operations 
and classical communication only 
\cite{Eisert02}. 
One promising way of realizing such non-Gaussian operations 
is to use measurement-induced nonlinearity, 
such as
photon subtraction from squeezed states 
with linear optics and photon counters 
\cite{Dakna97}.
The simplest such operation, which de-Gaussifies a single-mode squeezed 
state via a single photon subtraction, has recently been demonstrated 
\cite{Ourjoumtsev06,Neergaard-Nielsen06,NICT06}, 
where the photon subtraction was done using an on/off type photon detector, 
i.e. an avalanche photodiode.  

Extensions of this scheme to bipartite states 
enable one to perform entanglement distillation \cite{Browne03}. 
The primary concern is then 
to quantify the degree of entanglement enhancement. 
Methods for quantifying the entanglement of bipartite pure states 
have been established. 
In practice, however, imperfections are inevitable, 
making the output states mixed. 
For mixed states, especially non-Gaussian states, 
the evaluation of the entanglement is highly non-trivial. 
Two of the most frequently used measures of entanglement 
for mixed states are the negativity and logarithmic negativity (LN) 
\cite{Vidal02}. 
These are straightforwardly calculable even for the mixed states. 
The LN was used to predict the gain 
of the entanglement distillation with photon subtraction 
in practical settings with on/off detectors 
\cite{Kitagawa06}. 
Also, in the laboratory, the negativity 
of non-Gaussian entangled states has been evaluated 
by homodyne tomography 
for a particular type of photon-subtracted squeezed state 
where the entangled state is decomposable 
into a product of single-mode states, which greatly 
simplifies the tomographic process
\cite{Ourjoumtsev06-2}. 
In general, however, 
one needs a complete description of the state 
to compute the LN. 
It is a highly non-trivial task to reconstruct 
a multi-mode CV state via tomographic measurement 
without any a priori knowledge about the state 
(e.g. symmetry or decomposability of 
the state).

An indirect but practical approach is 
to examine the increase of the performance of concrete protocols 
in terms of their associated figures of merit 
such as the fidelity of teleportation \cite{NG_tlp}, 
degree of Bell inequality violation \cite{NG_Bell} 
and the mutual information of entanglement-assisted coding 
\cite{Kitagawa05}. 
These quantities can be evaluated using a homodyne-based measurement, 
such as the Bell measurement 
of the conjugate pair of quadratures $\hat{x} $ and $\hat{p} $ 
satisfying $[\hat{x},\hat{p}]=i/2$. 
This measurement is easily performed and does not require 
full tomographic reconstruction 
of the entangled state.

On the other hand, these quantities are not always true entanglement measures 
since they depend not only on the amount of entanglement 
but also on external factors, 
such as the input state for teleportation, 
the signal power and measurement strategy 
for entanglement-assisted coding. 
It might then be sensible to consider the limit of 
making these external factors as small as possible 
for assessing the intrinsic effect of entanglement. 
In particular, 
the close relationship between the LN and 
the mutual information of entanglement-assisted coding 
with weak modulation has been pointed out in 
\cite{Kitagawa06}. 
The mutual information 
in the limit of weak signal modulation 
is reduced to the Fisher information 
\cite{Braunstein}, 
which is a fundamental quantity in statistics 
\cite{Frieden}.

This implies another operational meaning of the LN, 
in addition to the one 
shown in \cite{Audenaert03}, 
i.e., 
the direct relationship between the LN and 
the entanglement cost 
under the positive-partial-transpose-preserving operations 
for certain classes of quantum states.  
The latter relies on the asymptotic manipulation scenarios, 
while the Fisher information can be directly accessed by 
a simple coding scheme.

In this paper, 
we discuss the relationship between the LN and the Fisher information 
of entanglement-assisted coding 
for important classes of CV states such as 
the two-mode squeezed states 
and the photon-subtracted squeezed states. 
We show that there is a direct connection 
between the LN and the Fisher information for these states. 
Moreover, a similar relationship is also found for some entangled qubit states.

\begin{figure}[b]
\centering 
\includegraphics[bb=55 55 345 340, width=.5\linewidth]{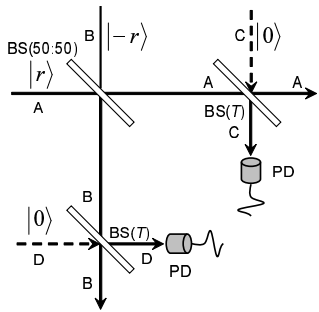}
\caption{\label{2mode-PS} Measurement-induced non-Gaussian operation 
on the two-mode squeezed state. }
\end{figure}

Let us denote the two-mode squeezed state as 
\begin{equation}
|r^{(2)} \rangle _{\rm AB} =\sum _{n=0} ^\infty \alpha _n 
|n\rangle _{\rm A} |n\rangle _{\rm B} , \label{SQ}
\end{equation}
where $\alpha _n \equiv \sqrt{1-\lambda ^2 } \lambda ^n $ 
with $0\leq \lambda =\tanh r<1$. 
The beam C (D) is tapped off from the beam A (B) 
with a beam splitter of transmittance $T\equiv 1-R$, 
\begin{equation}
|\psi \rangle _{\rm ABCD} =\hat{V} _{\rm AC} (\theta ) 
\hat{V} _{\rm BD} (\theta ) 
|r^{(2)} \rangle _{\rm AB} |0\rangle _{\rm CD}, 
\end{equation}
where 
$
\hat{V}_{ij} (\theta )
=\exp [\theta (\hat{a}_i ^\dagger \hat{a}_j 
-\hat{a}_i \hat{a}_j ^\dagger )] 
$
is the beam splitter operator, and 
$\tan \theta =\sqrt{(1-T)/T} $. 
Then the tapped beam C (D) is measured 
by a photon detector (Fig. \ref{2mode-PS}). 
\begin{eqnarray}
\hat{\rho } ^{\rm (AB)} 
=\frac{\textrm{Tr} _{\rm CD} [ 
|\psi \rangle _{\rm (ABCD)} \langle \psi |
(\hat{\Pi } _{\rm C} ^{\rm S} 
\otimes \hat{\Pi } _{\rm D} ^{\rm S} )] }{P_{\rm det} } , 
\end{eqnarray}
where $\hat{\Pi } _k ^{\rm S} $ 
is the POVM element corresponding to the event selection 
at the beam path $k$, and 
$
P_{\rm det} =\textrm{Tr} _{\rm ABCD} 
[|\psi \rangle _{\rm (ABCD)} \langle \psi |(\hat{\Pi } ^{\rm S} _{\rm C} 
\otimes \hat{\Pi } ^{\rm S} _{\rm D} )] 
$
is the success probability of the event selection. 
Assuming photon number resolving detectors, and when single photons 
are simultaneously detected at the beams C and D 
($\hat{\Pi } ^{\rm S} =|1\rangle \langle 1|$), 
the output state is projected onto the non-Gaussian pure state 
\begin{equation}
|\psi _{\rm NG} ^{(1)} \rangle _{\rm AB} 
=\frac{1}{\sqrt{P_{\rm det} ^{(1)} } } 
\sum _{n=0} ^\infty \alpha _{n+1} \xi _{(n+1),1} ^2 
|n\rangle _{\rm A} |n\rangle _{\rm B} , \label{NG}
\end{equation}
where 
$
\xi _{nk} \equiv (-1)^k \binom{n}{k} ^{1/2} 
(\sqrt{T} ) ^{n-k} ( \sqrt{R}) ^k , 
$
with the binomial coefficient $\binom{n}{k} $, and 
\begin{equation}
P_{\rm det} ^{(1)} =\frac{(1-\lambda ^2 )\lambda ^2 T^2 
(1+\lambda ^2 T^2 )}{(1-\lambda ^2 T^2 )^3 } 
\left( \frac{R}{T} \right) ^2 
\end{equation}

The LN 
\cite{Vidal02} 
of a quantum state $\hat{\rho } $ 
is defined by $E_\mathcal{N} =\log _2 ||\hat{\rho } ^{PT} ||$, 
where $||\cdot ||$ denotes the trace norm, and $PT$ means 
the partially transposition operation. For the pure state 
$|\chi \rangle _{\rm AB} 
=\sum _n c_n |n\rangle _{\rm A} |n\rangle _{\rm B} $, 
in particular, the LN is expressed as 
$E_\mathcal{N} (|\chi \rangle )=2\log _2 (\sum _n c_n )$. 
For the two-mode squeezed state (\ref{SQ}) 
and non-Gaussian pure state (\ref{NG}), 
\begin{eqnarray}
E_\mathcal{N} (|r^{(2)} \rangle )
&=&\log _2 \frac{1+\lambda }{1-\lambda } 
:= E_\mathcal{N} ^{\rm SQ} , 
\label{ln_SQ} \\
E_\mathcal{N} (|\psi _{\rm NG} ^{(1)} \rangle )
&=&\log _2 \frac{(1+\lambda T)^3 }{(1+\lambda ^2 T^2 )(1-\lambda T)} 
:= E_\mathcal{N} ^{\rm NG} , 
\label{ln_NG} 
\end{eqnarray}
respectively. In the range of 
\begin{equation}
0\leq \lambda \leq \lambda _{\rm LN} ^{\rm P} 
=\frac{-1+T+\sqrt{-7T^2 +18T-7} }{2T(2-T)} ,  
\end{equation}
$E_\mathcal{N} ^{\rm SQ} \leq E_\mathcal{N} ^{\rm NG} $ holds. 
In the limit as $T\rightarrow 1$, the non-Gaussian state 
$|\psi _{\rm NG} ^{(1)} \rangle $ always has higher entanglement, 
as quantified by the LN, than that of the two-mode squeezed state because 
$\lambda _{\rm LN} ^{\rm P} \rightarrow 1$.

\begin{figure}[b]
\centering 
\includegraphics[bb=0 0 1580 920, width=.9\linewidth]{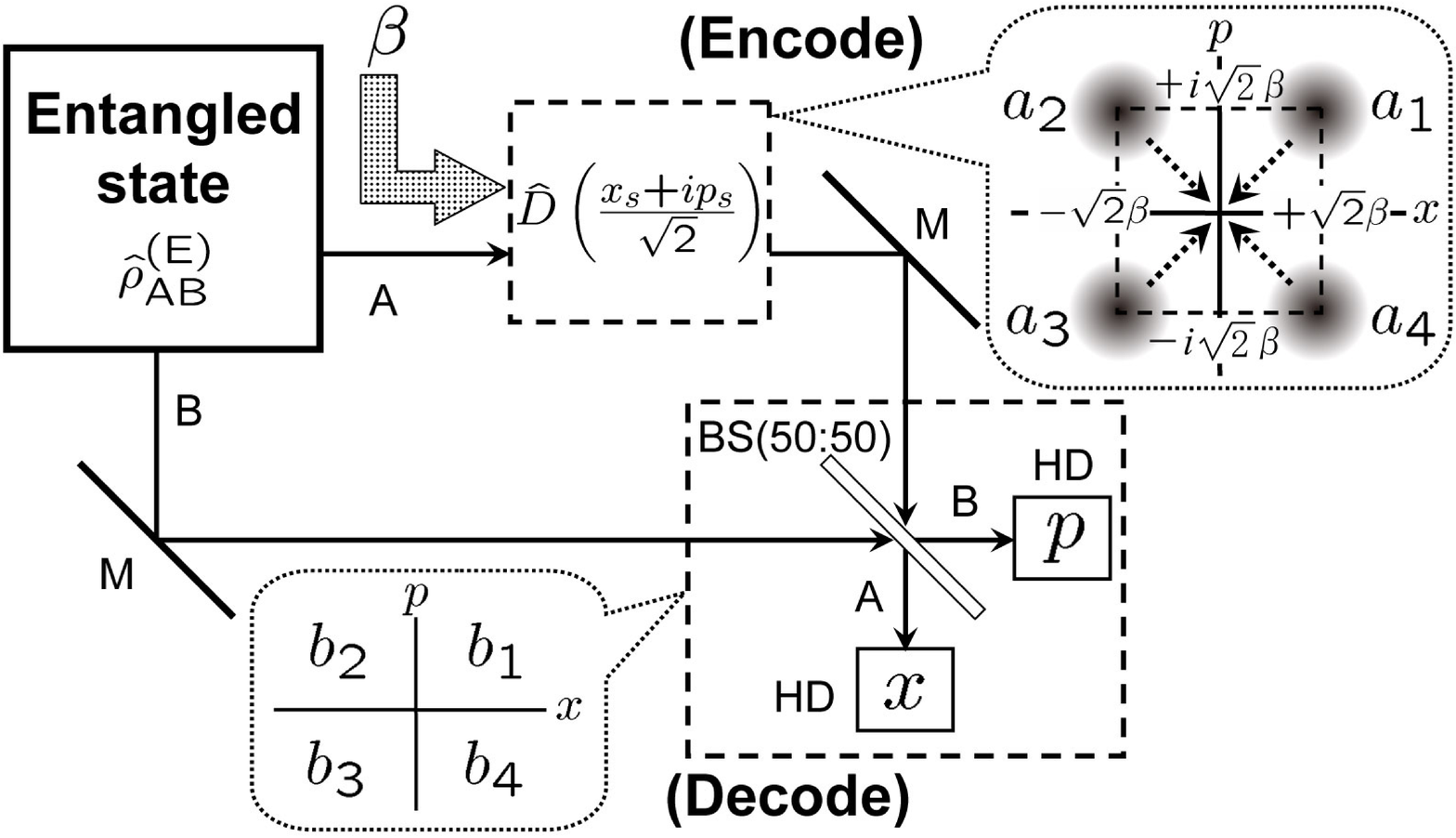}
\caption{\label{CV_coding} Entanglement-assisted coding channel of QPSK. }
\end{figure}

Let us now introduce the scheme for entanglement-assisted coding 
shown in Fig. \ref{CV_coding} 
\cite{Kitagawa05,Kitagawa06}. 
Initially, the sender Alice and the receiver Bob 
share an entangled state $\hat{\rho } ^{\rm (E)} _{\rm AB} $. 
Alice then encodes the quaternary phase-shift keyed (QPSK) signals 
$\{ a_k \}$ with equal likelihood $P(a_k )=1/4$ 
by the displacement operation on the beam A, 
$\hat{D}_{\rm A} [(x_s +ip_s )/\sqrt{2} ]$, 
with 
$x_s =\pm \sqrt{2} \beta $ 
and 
$p_s =\pm \sqrt{2} \beta $, 
where $\beta$ is the signal amplitude. 
Bob decodes the signals by the CV Bell measurement 
$|\Pi (x,p)\rangle _{\rm AB}$ 
and the quaternary decision on the measurement result $(x,p)$ 
into an appropriate quadrant $Q_l $ 
corresponding to the output $b_l$. 
The channel matrix is given by 
$P^{\rm (ch)} _{\beta ' =\beta } (b_l |a_k )
=\iint _{Q_l} dxdp P(x,p|x_s ^k ,p_s ^k )$, 
where 
$P(x,p|x_s ,p_s )={_{\rm AB} \langle } \Pi (x,p) |
\hat{U} _{\rm A} (x_s ,p_s )\hat{\rho } _{\rm AB} ^{\rm (E)} 
\hat{U} _{\rm A} ^\dagger (x_s ,p_s )|\Pi (x,p)\rangle _{\rm AB} $ 
is the conditional probability.

The mutual information is defined by 
\begin{equation}
I_\beta (\textrm{A;B})
=\sum _a P(a)\sum _b P^{\rm (ch)} _{\beta } (b|a) 
\log _2 \frac{P^{\rm (ch)} _{\beta } (b|a) }{P(b)} , 
\label{mutual info}
\end{equation}
where $P(b)=P^{\rm (ch)} _{\beta =0} (b|a)$. 
In the limit of small $\beta$ 
$P^{\rm (ch)} _{\beta } (b|a)$ and 
$\log _2 P^{\rm (ch)} _{\beta } (b|a)$ 
can be expanded 
in a series around $\beta =0$ as 
\begin{eqnarray}
\lefteqn{I_\beta (\textrm{A;B})} \nonumber \\
&&=\sum _a P(a)\sum _b \frac{\beta ^2 }{2\ln 2} 
\frac{1}{P_0 ^{\rm (ch)} (b|a)} 
\left[ \frac{dP_{\beta ' } ^{\rm (ch)} (b|a)}{d\beta '} 
\Bigg|_{\beta ' =0} \right] ^2 \nonumber \\
&&\hspace{60mm} +O(\beta ^3 ). 
\end{eqnarray}
Using the Fisher information for a parameter $\beta ' $ 
\cite{Frieden}
\begin{equation}
J_{\beta ' } =\sum _b \frac{1}{P_{\beta ' } ^{\rm (ch)} (b|a)} 
\left\{ \frac{dP_{\beta ' } ^{\rm (ch)} (b|a)}{d\beta ' } \right\} ^2 , 
\end{equation}
and taking the symmetry of the channel into account, 
we obtain 
\begin{equation}
\lim _{\beta \rightarrow 0} 
\frac{I_\beta (\textrm{A;B}) }{\beta ^2 } 
=\frac{J_0 }{2\ln 2} , 
\end{equation}

\begin{figure}
\centering 
\includegraphics[bb=65 65 560 750, angle=-90, width=.9\linewidth ]{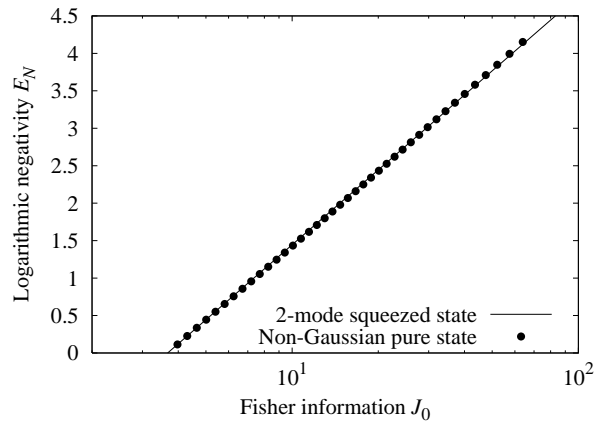}
\caption{\label{Fi-ln_pure} 
Correlation between the Fisher information 
and logarithmic negativity for the two-mode squeezed states (solid line) 
and the photon-subtracted squeezed states (points) 
in various levels of a parameter $\lambda T$ (the combination of squeezing degree and 
beam splitter transmittance). }
\end{figure}

For the two-mode squeezed state (\ref{SQ}) 
and the non-Gaussian pure state (\ref{NG}), 
the Fisher informations are 
\begin{eqnarray}
J_0 ^{\rm SQ} &=&\frac{8}{\pi \ln 2} \frac{1+\lambda }{1-\lambda } . 
\label{Fi_SQ} \\
J^{\rm NG} _0 &=&\frac{1}{2\pi \ln 2}\frac{(3\lambda ^2 T^2 +4\lambda T+4)^2 }
{(1+\lambda ^2 T^2 )^2 }
\frac{1+\lambda T}{1-\lambda T} , \label{Fi_NG}
\end{eqnarray}
respectively. With Eqs. (\ref{ln_SQ}) and (\ref{ln_NG}), we obtain 
\begin{eqnarray}
E_\mathcal{N} ^{\rm SQ} &=&\log _2 \frac{\pi \ln 2}{8} 
J^{\rm SQ} _0 , \label{Fi-ln_SQ} \\
E_\mathcal{N} ^{\rm NG} &=&\log _2 f(\lambda T)\frac{\pi \ln 2 }{8} 
J^{\rm NG} _0 . \label{Fi-ln_NG}
\end{eqnarray}
where 
\begin{equation}
f(\lambda T)=\frac{16(1+\lambda T)^2 (1+\lambda ^2 T^2 )}
{(3\lambda ^2 T^2 +4\lambda T+4)^2 } , 
\end{equation}
which is very close to unity for $0\leq \lambda T<1$. 
In Fig. \ref{Fi-ln_pure}, 
the correlation 
between $J_0 ^{\rm NG} $ and $E_\mathcal{N} ^{\rm NG} $ 
for various $\lambda T$ is indicated. 
In particular, 
for $0\leq \lambda T\lesssim 0.8$, the difference of $f(\lambda T)$ 
from unity is within $\pm 2\%$, 
showing that Eq. (\ref{Fi-ln_SQ}) and Eq. (\ref{Fi-ln_NG}) 
are almost identical relations 
in a good approximation. 
Thus the Fisher information of the entanglement-assisted coding 
has monotonic correspondence with the LN 
for the two-mode squeezed states and 
the photon-subtracted squeezed states.

\begin{figure}
\centering 
\includegraphics[bb=60 65 555 750, angle=-90, width=.9\linewidth]{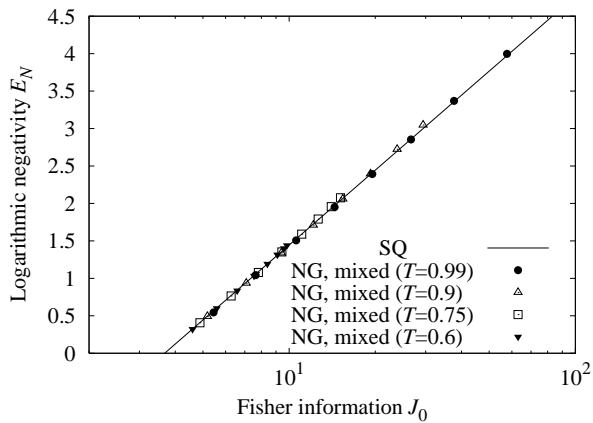}
\caption{\label{Fi-ln_mixed} Correlation between the Fisher information 
and logarithmic negativity for the two-mode squeezed state (solid line) 
and non-Gaussian mixed states of photon-subtracted squeezed state 
(variety of points). }
\end{figure}

We then consider a more practical situation, 
where the photon subtraction is made by 
on/off type photon detectors, which is described by the set of POVM elements
$\{\hat{\Pi } ^{\rm (off)} =|0\rangle \langle 0|, 
\hat{\Pi } ^{\rm (on)} =\hat{1}-|0\rangle \langle 0| \}$. 
In this case the generated state conditioned by the `on' 
signal is generally a mixed state, as described by 
\begin{equation}
\hat{\varrho } _{\rm NG} =\frac{1}{\mathcal{P} _{\rm det} } 
\sum _{i,j} |\Phi _{ij} \rangle _{\rm (AB)} \langle \Phi _{ij} |, 
\end{equation}
where 
$|\Phi _{ij} \rangle _{\rm AB} 
=\sum _{n=\max \{i,j\} } ^\infty \alpha _n \xi _{ni} \xi _{nj} 
|n-i\rangle _{\rm A} |n-j\rangle _{\rm B} $ and 
\begin{equation}
\mathcal{P}_{\rm det} =\frac{\lambda ^2 (1-T)^2 (1+\lambda ^2 T)}
{(1-\lambda ^2 T)(1-\lambda ^2 T^2 )} . 
\end{equation}
As the transmittance of tapping beam splitter get smaller, 
the mixedness of output state increases. 
The figure \ref{Fi-ln_mixed} illustrates the relationship between the Fisher information 
and LN for the photon-subtracted mixed states with various values of the transmittance. 
Here, the LN is calculated numerically using the methods described in \cite{Kitagawa06}. 
We see here that, 
even for the mixed state cases, the same relation as Eq. (\ref{Fi-ln_SQ}) 
holds in a fairly good approximation. 
The accuracy of this approximation 
is within $\pm 2\%$ for $0\leq \lambda T\lesssim 0.8$, 
which is sufficient to verify the entanglement enhancement 
via the measurement-induced non-Gaussian operation. 
For $\lambda =0.4$, 
for example, 
the Fisher information with the two-mode squeezed state is 
$J_0 ^{\rm SQ} =8.572$, 
and those of non-Gaussian pure and mixed states 
of photon-subtracted squeezed states 
with $T=0.9$, are $J_0 ^{\rm NG(P)} =12.992$ 
and $J_0 ^{\rm NG(M)} =12.153$, respectively. 
The degrees of improvement are 51.6\% and 41.8\%, 
respectively, 
which are much greater than the accuracy of the approximation.

We finally show that 
the entanglement evaluation with the Fisher information 
is applicable to other entangled states. 
An interesting example is entangled photon-number-qubit state 
such as $\hat{\rho }_{\rm qubit} =t|\xi \rangle \langle \xi | 
+(1-t)[|c_0 |^2 |0\rangle \langle 0|\otimes |1\rangle \langle 1| 
+|c_1 |^2 |1\rangle \langle 1|\otimes |0\rangle \langle 0|]$, 
where 
$|\xi \rangle =c_0 |0\rangle |1\rangle +c_1 |1\rangle |0\rangle $ 
and the parameter $0\leq t\leq 1$ 
indicates its mixedness. 
The coefficients satisfy $|c_0 |^2 +|c_1 |^2 =1$ 
and $c_1 $ has the relative argument $\phi $ to $c_0$. 
Applying such a state to our entanglement-assisted coding scheme, 
the conditional probability $P_{\rm qubit} (x,p|x_s ,p_s)$ is obtained 
as a function of $t|c_0 ||c_1 |$ and $\phi $, 
and so is the Fisher information 
$J_0 ^{\rm qubit} $. 
The dependency on $\phi $ is caused 
by our specific choice of QPSK encoding, 
and averaging $J_0 ^{\rm qubit} $ with respect to $\phi $ removes 
its arbitrariness. 
The average Fisher information $\bar{J}_0 ^{\rm qubit} 
=\frac{1}{2\pi } \int _0 ^{2\pi } d\phi J_0 ^{\rm qubit} $ 
is now a function of the unit $t|c_0 ||c_1 |$ only. 
On the other hand, 
the LN for $\hat{\rho } _{\rm qubit} $ is evaluated 
with $E_\mathcal{N} ^{\rm qubit} =\log _2 (1+2t|c_0 ||c_1 |)$. 
Therefore, 
it is also the case here that 
$\bar{J}_0 ^{\rm qubit} $ and $E_\mathcal{N} ^{\rm qubit} $ 
are directly related, 
although this is a different relationship from that 
expressed by Eqs. (\ref{Fi-ln_SQ}) and (\ref{Fi-ln_NG}).

For the other entangled photon-number-qubit states 
such as 
$|\xi ' \rangle =c_0 ' |0\rangle |0\rangle +c_1 ' |1\rangle |1\rangle $ 
and their associated mixed states, 
one cannot find the one-to-one correspondence 
between the Fisher information and the LN any longer in our scheme. 
However, $|\xi ' \rangle $ can be transformed 
into $|\xi \rangle $ with the local flipping of $0\leftrightarrow1$. 
This operation can, with current technology, 
be performed unitarily to arbitrarily high accuracy 
and thus in a way which preserves entanglement \cite{Kitagawa03, Berry06}. 
This means that, by modifying the decoding strategy, 
inserting the flipping operation, 
the Fisher information can turn to work again.  
These two examples imply that, 
under a suitable encoding and decoding strategy, 
the correspondence between the Fisher information and the LN 
might always appear for arbitrary bipartite entangled state.

In this paper, we showed a direct relationship between 
the logarithmic negativity and the Fisher information 
of the entanglement-assisted coding 
for important classes of CV states, i.e. 
the two-mode squeezed states 
and the photon-subtracted squeezed states, 
and also entangled qubit states.  
This is the first observation of direct connections 
between the logarithmic negativity and 
the Fisher information, 
and we have found that they hold 
for a wide class of bipartite entangled states. 
It is an important future goal to generalize this 
to a wider class of states by optimizing 
the encoding and decoding strategies.

The authors would like to acknowledge S.~L.~Braunstein, M.~Ban, 
J.~A.~Vaccaro, and B.~C.~Sanders for their valuable discussions. 
AC was supported by the EU project QAP.

\end{document}